\documentclass[final,3p,times,twocolumn]{elsarticle}
\usepackage{etoolbox}
\makeatletter
\patchcmd{\ps@pprintTitle}{\footnotesize\itshape
       Preprint submitted to \ifx\@journal\@empty Elsevier
       \else\@journal\fi\hfill\today}{\relax}{}{}
\makeatother
\pdfoutput=1

\usepackage{blindtext, graphicx, amsmath, algorithm, algpseudocode, pifont, algcompatible, comment, layout, amsthm, amssymb}
\usepackage{enumitem}   

\usepackage{float}

\usepackage[utf8]{inputenc}
\usepackage[english]{babel}

\begin{document}

\begin{frontmatter}

\title{A Brief Survey on Autonomous Vehicle Possible Attacks,Exploits and Vulnerabilities}
\author{Amara Dinesh Kumar\corref{cor1}}
\ead{dineshkumar.amara@gmail.com}
\author{Koti Naga Renu Chebrolu}
\address{Department of Electronics and Communication Engineering, Amrita School of Engineering, Coimbatore,\\ Amrita Vishwa Vidyapeetham, India}
\author{Vinayakumar R\corref{cor1}}
\author{Soman KP}
\address{Center for Computational Engineering and Networking (CEN), Amrita School of Engineering, Coimbatore,\\ Amrita Vishwa Vidyapeetham, India}

\begin{abstract}
 Advanced driver assistance systems are advancing at a rapid pace and all major companies started investing in
developing the autonomous vehicles. But the security and reliability is still uncertain and debatable. Imagine that a vehicle is compromised by the attackers and then what they can do. An attacker can control brake, accelerate
and even steering which can lead to catastrophic consequences. This paper gives a very short and brief overview of most of the possible attacks on autonomous vehicle software and hardware and their potential implications.
\end{abstract}


\end{frontmatter}


\vspace{-0.4cm}
\section{Introduction}
\label{sec1}
Digital technologies are improvising at a rapid speed and moving towards automation by including artificial intelligence. Similarly Automakers also started focusing on connected Autonomous vehicles. These vehicles help to improve the safety of passengers and increase the efficiency of transportation by interacting with the external world through V2X and V2V communications. By sharing data like speed, position, and heading angle help to predict the succeeding location of the vehicle. Using Sensor technologies, maneuvering of roads and terrains becoming easier and with the help of predetermined knowledge\cite{8303906}.

As the percentage of automation increases the human interaction with the vehicle increases and passengers will become the audience. This may also lead to many potential cyber threats and attacks like Hacking ECU's, GPS spoofing, Modified traffic signs, Injecting false bits in the CAN and altering sensor values. This paper very briefly discuss about the  vulnerabilities, security issues, exploitative methods and the adverse effect of them on the connected autonomous vehicles\cite{6504448}. 
\vspace{-0.3cm}
\subsection{Why are autonomous vehicles more vulnerable to attacks?}
Compared to conventional vehicle autonomous vehicle has to communicate with other vehicles and infrastructure which are external networks and may become a channel for attack and an opportunity for hackers\cite{8328277}.
For autonomous vehicles, vast information has to be processed particularly deep learning algorithms are used in processing image related data which are prone to adversarial attacks and the probability of false positives increases with increase in the amount of data to be processed.
As the technology is in the nascent stage in both hardware and software and not yet rigorously tested because of which it is difficult to make it reliable in all conditions and also autonomous vehicles are prone to bugs and vulnerabilities also which can be exploited by the attackers\cite{7413993}.

\begin{figure*}[h]
   \centering
   \label{fig1}
   \includegraphics[width =18cm,height =6cm]{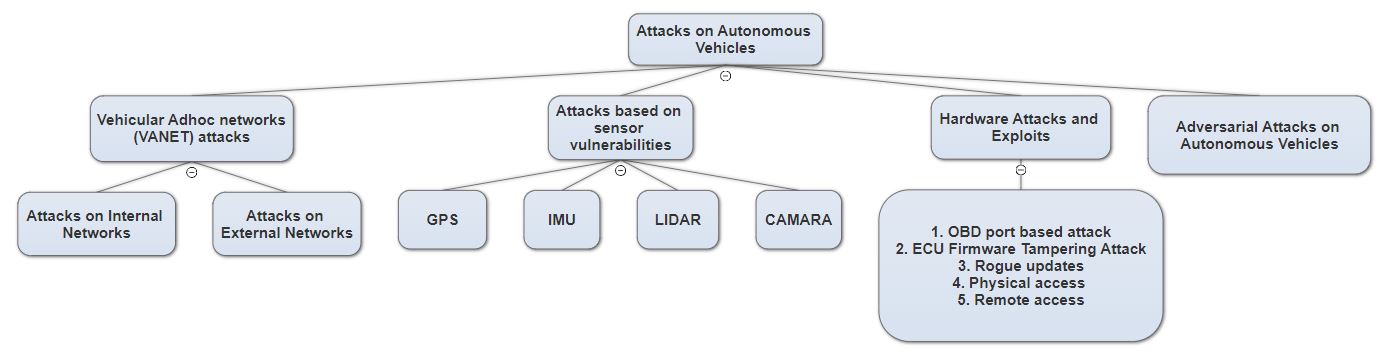}
   \caption{Possible attacks on autonomous vehicles}
\end{figure*}
\vspace{-0.4cm}
\section{Attacks exploiting the vulnerabilities in various sensors of vehicle}
\label{sec2}
\vspace{-0.2cm}
\subsection{Global Positioning System (GPS)}
To locate and navigate the vehicle, one uses GPS data with great accuracy. To overcome the difficulties in getting GPS data, the count of satellites increased in the public domain where one can easily access the data. The provision of free access to the data with transparent architecture helping the hackers to mislead or manipulate the data to provide the wrong directions or to control the routing of the vehicle\cite{8126120}. This lead to the security and safety issues of the passengers. Misleading of the GPS signals is known as GPS spoofing and jamming where hackers transmit the unrealistic signal/data. The strength of the unrealistic signal increases as the GPS receivers are programmed to receive the stronger signals and gradually the position of the vehicle is modified from the desired target\cite{7872388}.
\vspace{-0.3cm}
\subsection{Inertial Measurement Unit (IMU)}
IMU is the combination of the Gyroscope and Accelerometers which provides the data of velocity, acceleration, and orientation of the vehicle. They also monitor the change in the environmental dynamics like the gradient. The data provided by the sensors can be modified or not to recognize the gradient of the road. It causes the vehicle to moves slow on the gradient roads which intern slow down the following vehicles\cite{7872388}.
\vspace{-0.3cm}
\subsection{Light Detection and Ranging (Lidar)}
Light Detection and Ranging (LiDAR) is used to localize the environment, obstacle detection, and avoidance. It based on the technology,  time taken by the light to travel to and fro to the vehicle determine's the distance at which the object is located. If the hacker sends a signal of the same frequency to the scanner and to assume the object is detected. It makes the Autonomous vehicle to move slowly or stop\cite{7872388}.
\vspace{-0.3cm}
\subsection{Monoscopic and Stereoscopic Cameras}
Cameras are used to detect lane detection, traffic sign recognition, headlight detection, obstacle detection, etc. Functioning of cameras can be partially disabled by using high beam torches or headlights of the opposite vehicles. It may introduce the safety concerns like false detection or not detection of the objects. Complementary metal oxide(CMOS) sensors used in the camera can be blinded by the high-power lights\cite{7872388}.
\vspace{-0.4cm}
\section{Vehicular Adhoc Network (VANET) attacks}
\label{sec2}
\vspace{-0.3cm}
\subsection{Attacks on Internal Networks}

\subsubsection{Passcode and Key attacks}
passcode and keys are one of the safety features of the connected cars. Multiple attempts may crack the password or keys which works on the IR based technologies can be easily hacked. Brute force attack on the passcode can crack the passcodes of the connected cars. Bluetooth connectivity can be affected and leaks the private data of the passenger.
\vspace{-0.3cm}
\subsection{Attacks on External Networks}

\subsubsection{V2X Network Attacks}
Connecting car concept and emerging technologies in vehicular networks and which can communicate with other cars and infrastructure giving many advantages along with opening up the network and providing the network access point for the attackers to exploit\cite{7223297}. It can connect to a smartphone,cloud and other devices and communicate which is described as the V2X communication.

Generally communicating channel between the car and smartphone is established through Wifi, Bluetooth and GSM protocols which are inherently vulnerable and contain known bugs and vulnerabilities which can be exploited by the attackers. Connecting to a smartphone is always a risk for a vehicle as it is interacting with an external unfamiliar device. Sending and receiving data from the cloud is a threat as the data centre may get compromised and then vehicle starts communicating with the miscellaneous server.

Dedicated Short Range Communication (DSRC) protocol in V2V networks is a duplex communication protocol channel used particularly for automotive use operating at 5.9 GHz with a bandwidth of 75 Mhz, WAVE (Wireless Access in vehicular environments) and IEEE 802.11 p are the protocols that are generally used in the V2X communication. All of them have known vulnerabilities which can be exploited by an attacker.

Vehicle to Vehicle (V2V) Network Attacks: The Communication between the host vehicle and adjacent vehicles for overtaking, lane changing, at intersection cars, exchange data through V2V networks.
Impersonation attack consists of a malicious car which connects with the host vehicle with a false identification by spoofing then it establishes communication sending the malicious and receiving the sensitive data capturing it than logging and storing it\cite{8071577}. 

The major drawback of V2V communication is the use of insecure and unencrypted protocols which makes attackers to eavesdropping the traffic and data between the host vehicle and other vehicle communication and then get the sensitive information like authentication keys leading to authentication attacks 

Vehicle to Infrastructure (V2I) Network Attacks: The vehicle connects with the infrastructure establishing a communication channel for receiving and transmitting information. The vehicle connects with intelligent traffic signs and cellular network nodes which can be compromised, infected and impersonated by an attacker thus gaining a access through a backdoor intruding into the vehicle network and ECU's.

\vspace{-0.3cm}
\subsection{Distrubuted Denial Of Services (DDOS)}
Denial of service is one of the most vulnerable threats that connected cars may experience. service of the system denied by the several attacking mechanisms which results in disruption of the traffic flow and damages to the infrastructure. It may cause the collision of the vehicles, a life threat for the passengers.
\vspace{-0.4cm}
\section{Hardware attacks and exploits}
\vspace{-0.3cm}
\subsection{OBD port based attack}
OBD stands for onboard diagnostics and OBD port is present in almost all the vehicles manufactured from 2008\cite{checkoway2011comprehensive}. OBD port is used for collecting the diagnostics data of the vehicle. It gives the data about the vehicle faults and performance. It interacts with the ECU's communicating through CAN bus. It is a hand held device like USB which has to be connected to the vehicle through the port generally present below the dashboard opposite to adjacent driver seat. which then connects to the computer through a wired connection using USB port or through a wireless connection using Bluetooth\cite{zhang2016controlling}. once connected PC can send and receive the data to and from the vehicle ECU's and with a possible exploitation can also manipulate the data packets and inject malicious packets into the vehicle network\cite{yan2015two}. 
\vspace{-0.3cm}
\subsection{ECU Firmware tampering attack}
ECU (Engine control unit) is a electronic control module for the sensors and actuators of any sub-system in a vehicle and a typical vehicle consists of more than 100 ECU's\cite{7908939}.
ECU code is propitiatory making it safe and secure but attackers started targeting recent attacks by re flashing the ECU with custom firmware altering its state and inducing malicious and unintended actions.
It is called a direct access attack as we assume that attacker has physical access to the ECU.Attacker updates the firmware of ECU using the external interface thus altering the functionality of ECU.By altering the ECU memory and tampering the security keys and maintaining the integrity of the ECU firmware code and its updates using the hashing techniques and authentication for software updating\cite{7908939}.

\vspace{-0.3cm}
\subsection{Rogue updates}

Firmware updates in the connected cars are one of the sources for the Rogue updates. These updates are not from the manufacturer and lack of proper updates with safety and security. they are prone to the severe cyber attacks which leak private data of the vehicle. It allows hackers to provide enough security weakness to introduce malware and control the firmware of the connected cars.

The exploitation can be through 1) Physical Access 2) Remote Access

\subsubsection{Physical Access}
Nowadays physical layer is integrated with the ECU's directly which increases the potential for cyber attacks. Hackers can directly exploit the sensor data, control and communication modules. The attacks can be direct or indirect attacks targeting vehicular electronic modules or overloading effect on the physical layer.

\subsubsection{Remote Access}
Remote Access can be done through different connections like WiFi, Bluetooth, 4G, etc. ECU's are directly connected to the CAN bus which is not intended. Connectivity to internet leads to the potential threat of injecting malware or virus files into the firmware. Automakers are not sure about the threats or actions to be taken to eradicate them. Even Automakers are not sure of the recovery actions to be performed.
\vspace{-0.4cm}
\section{Adversarial Attacks on Autonomous vehicles}

Deep Learning is based on the probabilistic estimation for the classification tasks predicting the data belonging to a particular class with a certain level of probability and when the probability is higher then we say that class has higher confidence\cite{kumar2018novel}.

Deep learning algorithms like Deep Neural Networks (DNN's) are start of the art techniques that are used in the autonomous vehicles for perception and sensing\cite{8126154}. But DNN's are prone to certain adversarial attacks making the autonomous vehicles vulnerable to such attacks.Adversarial attack is a form of attack in which the small noise or perturbations are added to the original data misleading and deceiving the DNN's decision making which may create dangerous consequences\cite{DBLP:journals/corr/abs-1801-00553}.

In \cite{DBLP:journals/corr/EvtimovEFKLPRS17} physical perturbations are created by adding stickers to a stop traffic sign at certain positions and the traffic sign detection algorithm failed to classify it correctly.

\vspace{-0.4cm}
\section{Miscellaneous Attacks}
Numerous attacks were getting implemented by many researchers and hackers.Gaining access to key less entry system by spoofing,Malware attacks and applications that installed in the infotainment and navigation systems. New types of attacks like attacking the electric vehicle by exploiting the electric charger which is connecting and logging the data.

Because of the diversity in the automobiles like combustion engine cars,hybrid cars,plug in hybrid and complete electric cars along with the autonomous vehicle capabilities it is difficult to maintains the overall security and presence of different OEM's(original equipment manufacturer) and tier 1 manufacturers which makes a very complex supply chain for the automobiles thus creating the scope for many miscellaneous vulnerabilities and exploitation's.
\vspace{-0.4cm}
\section{Conclusion}
This paper aims to cover most of the common cyber attacks and exploits that are possible on a autonomous connected vehicle. This is most dynamic area and the attacks are getting sophisticated day by day and attackers are always finding new ways and tools to deceive and hack the vehicle. As the autonomous vehicle development is growing with rapid pace but the security aspect of vehicle is not receiving deserved attention which may become a serious threat to autonomous vehicles security and adoption as many countries are trying to bring autonomous vehicles on road soon. Researchers have to come forward to join hands for collaborating and proactively giving priority to cyber security at design and development stages. 

\bibliographystyle{elsarticle-num}

\vspace{-0.4cm}
\bibliographystyle{unsrt}
\bibliography{ref}
\end{document}